\documentclass{elsart}
\usepackage{amsfonts}
\usepackage{amsmath}
\usepackage{graphicx}
\usepackage{amssymb}

\def\comment#1{}

\def\mn#1{*\marginpar{*\tiny{#1}}}
\def\mn#1{}

\usepackage{color}
\usepackage{ulem}
\usepackage{cancel}
\def\comment#1{}
\def\red#1{\textcolor{red}{#1}}
\def\blu#1{\textcolor{blue}{#1}}

\def\mn#1{\marginpar[\tiny{\red{#1}}]{\tiny{\blu{#1}}}}

\begin{document}
\begin{frontmatter}
\title{Gravitational and electric energies in the collapse of a spherical thin-shell capacitor}
\author
{Remo Ruffini and She-Sheng Xue}
\address
{ICRANet Piazzale della Repubblica, 10-65122, Pescara, \\and Physics Department, University of Rome "La Sapienza," P.le A. Moro 5, 00185 Rome, Italy}

\begin{abstract}
We adopt a simplified model describing the collapse of a spherical thin-shell capacitor to give an analytical description how gravitational energy is converted to both kinetic and electric energies in the gravitational collapse. It is shown that (i) averaged kinetic and electric energies are the same order, about an half of gravitational energy of spherical thin-shell capacitor in the collapse; (ii) caused by radiating and rebuilding electric energy, the gravitational collapse undergoes a sequence of ``on and off'' hopping steps in the microscopic Compton scale. Although the collapse process is still continuous in terms of macroscopic scales, it is slowed down as the kinetic energy is reduced and collapsing time is about an order of magnitude larger than that of the collapse process eliminating electric processes. 
\end{abstract}

\end{frontmatter}

\noindent {Email}: xue@icra.it, Tel. 003908523054204 and Fax: 00390854219252 

\noindent
{\bf Introduction.}
\hskip0.1cm
In the gravitational collapse of neutral stellar cores at densities comparable to the nuclear density, both macroscopic processes of gravitational and hydrodynamical interactions and microscopic processes of the strong and electroweak interactions occur. In theoretical principle, these can be well described by the Einstein-Maxwell equations and the equations for the number and energy-momentum conservation of particles, duly taking into account their interactions. In practical calculations of analytical or numerical approach, however, it is rather difficult to simultaneously analyze both macroscopic and microscopic processes for the reason that the time and length scales of macroscopic processes are much larger than those of the microscopic processes.   
The approximation normally adopted is that microscopic processes are treated as local and instantaneous processes which are effectively represented by a model-dependent parameterized equation of state (EOS). We call this {\it approximate locality}. 

Applying the {\it approximate locality} to electric processes, as required by the charge conservation, one is led to {\it local neutrality}: positive and negative charge densities are exactly equal overall space and time. As a consequence, all electric processes are completely eliminated in the assumption of the {\it approximate locality}. On the other hand, it is well known that an internal electric field (charge-separation) {\it must} be developed \cite{ob_1975} in a totally neutral system of proton and electron fluids in the presence of gravitational fields. If the electric field (process) is weak (slow) enough, the {\it approximate locality} is applicable. However, this 
should be seriously questioned when the electric field (process) is strong (rapid) in the case that neutral stellar cores reach the nuclear density where positive charged baryons interact via strong interactions that do not associate to negative charged electrons, in addition to widely different gravitational masses of baryons and electrons. In fact, strong electric fields are created on the baryon core surface in an electrostatic equilibrium state \cite{Usov1}.
Furthermore, it is shown in Ref.~\cite{hrx2012}, either pulsating or gravitationally collapsing of the baryon core results in the dynamical evolution of electrons, as a consequence, the strong electric field dynamically evolves in space and time, and leads to the electron-positron pair-production process of Sauter-Heisenberg-Euler-Schwinger (see the review \cite{phreport}) for overcritical electric fields $E\gtrsim E_c\equiv m_e^2c^3/(e\hbar)$. 
When this occurs in gravitational collapses of neutral stellar cores, some part of the gravitational energy of neutral stellar cores converts to the observable energy of electron-positron pairs, as a result, the kinetic and internal energies of neutral stellar cores are reduced. 

As mentioned above, the difficulties of dealing with such a problem come from very different space-time scales of macroscopic and microscopic processes. We are forced to properly split the problem into three parts: (i)
microscopic processes of electrodynamics; (ii) macroscopic processes of gravitational collapses; (iii) the back-reaction of microscopic processes on macroscopic processes. In Ref.~\cite{hrx2012}, we study the first part of the problem: microscopic processes of electrodynamics for strong electric field oscillations and pair-productions, which form a radiative electric energy, in a postulated space-time world line of gravitational collapse. However, the back-reaction of such radiative electric energy on collapse was not considered.       
In this article, we start to quantitatively understand the second and third parts of the problem in a simplified model how gravitational, electric and kinetic energies of neutral stellar cores transfer from one to another in gravitational collapses, to see the possibility of converting the gravitational energy to the electromagnetic energy by the ``breaking process'' of reducing kinetic energy \cite{RV2003}. 
The Planck units $G=\hbar=c=1$ are adopted, unless otherwise specified.
  
\noindent
{\bf Einstein-Maxwell Equations and conservation laws of two fluids}
\hskip0.1cm
The gravitational collapse of neutral
stellar cores is generally described by the Einstein-Maxwell equations and those governing the particle number and
energy-momentum conservation 
\begin{align}
G_{\mu\nu} &=  - 8 \pi G (T_{\mu\nu}+T^{\rm em}_{\mu\nu}),\quad F^{\mu\nu}_{\,\,\,\,\,\,\,;\nu} = 4\pi J^{\mu},\nonumber\\
(T^{\nu}_{\,\,\,\mu})_{;\nu} &= -F_{\mu\nu}J^{\nu},\quad
\quad\quad\,\,\,\,\,\,\,\,  (\bar n_{e,B}U^\nu_{e,B})_{;\nu}=0,\label{coeqns1}
\end{align}
in which appear the Einstein tensor $G_{\mu\nu}$,  the electromagnetic field $F^{\mu\nu}$ (satisfying $F_{[\alpha \beta, \gamma]}=0$) and its energy-momentum tensor
\begin{align}
T^{\rm em}_{\mu\nu}=\frac{1}{4\pi}\left(F_\mu^{\,\,\rho} F_{\rho\nu}-\frac{1}{4}g_{\mu\nu}F^{\rho\sigma} F_{\rho\sigma}\right);
\label{tem}
\end{align}
$U^\nu_{e,B}$ and $\bar n_{e,B}$ are respectively the four-velocities and proper number-densities of electrons and baryons, 
\begin{equation}
J^\mu= e\bar n_pU^\mu_{B}-e\bar n_eU^\mu_{e}
\label{ecurrent}
\end{equation} 
is the electric current density, and $\bar n_p < \bar n_B$ the proper number-density of the positively charged baryons.
The energy-momentum tensor $T^{\mu\nu}=T_e^{\mu\nu}+T_B^{\mu\nu}$ is taken to be that 
of two simple perfect fluids representing electrons and the baryons, 
each of the form
\begin{align}
T^{\mu\nu}_{_B} &= \bar p_{\!_B}g^{\mu\nu}+(\bar p_{\!_B}+\bar\rho_{\!_B})U^\mu_{\!_B} U^\nu_{\!_B},
\label{etensor}\\
T^{\mu\nu}_{e} &= \bar p_{e}g^{\mu\nu}+(\bar p_{e}+\bar\rho_{e})U^\mu_{e} U^\nu_{e},
\label{etensor_e}
\end{align}
where $\bar\rho_{_{e,B}}$ and $\bar p_{_{e,B}}$ are the respective proper energy densities and pressures. In this scenario, electrons and baryons are respectively described by two perfect fluids at or over the nuclear density, and they couple each other via the electromagnetic interaction.  

Baryon fluid and electron fluid must be separately described for the reasons that in addition to the different kinematics of baryons and electrons, the most important differences between their dynamics are: (i) baryons are much more massive than electrons in terms of the long-range gravitational force and baryon cores undergo relativistically collapsing processes; (ii) at or over the nuclear density $\bar n_{\rm nucl}$, the electron pressure is much larger that baryon one, and baryons interact each other via the short-range strong force that does not act on electrons. Electron and baryon fluids interact via the long-range electromagnetic force, when two fluids are at or over the nuclear density, this interaction between two fluids becomes rather strong, as will be specified below. Note that we ignore the short-range weak interactions for the $\beta$-process in this article. The long-range gravitational and electromagnetic forces are explicitly present in Eqs.~(\ref{coeqns1}-\ref{ecurrent}). Instead, the short-range strong interaction is taken into account by pressure and energy density in the proper frame (see Ref.~\cite{Weinberg1972}),
\begin{align}
\bar p_{\!_B} &=\frac{1}{3} \sum_{i=1}^3 T^{ii} = \frac{1}{3} \sum_{_B} \delta^3({\bf x}-{\bf x}_{_B})\frac{{\bf p}_{_B}^2}{E_{_B}},\label{p_b}\\
\bar \rho_{\!_B} &= T^{tt} = \sum_{_B} \delta^3({\bf x}-{\bf x}_{_B})E_{_B} 
\label{e_b}
\end{align}
where $E_{\!_B}=E_{\!_B}({\bf p}_{\!_B})$ is the energy spectrum of baryons, duly taking into account their short-range strong interactions (nuclear potential) at a given density $\bar n_{_B} \gtrsim \bar n_{\rm nucl}$. Electrons' pressure and energy density are analogously given by Eqs.~(\ref{p_b}) and (\ref{e_b}) by replacing the subscript $B\rightarrow e$, however, the spectrum $E_{e}=E_{e}({\bf p}_{e})$ is different from baryon one, due to the fact that electrons are blind with the short-range strong interactions. As a result, the baryon and electron EOS $\bar p_B=\bar p_B(\bar\rho_B)$ and $\bar p_e=\bar p_e(\bar\rho_e)$ are different, moreover, the space-time gradients $\nabla \bar p_{_{e,B}}$  and $\partial \bar p_{_{e,B}}/\partial t$ are different. 

Now we turn to discuss how the short-range strong interaction effect on the baryon fluid velocity $v^i_{_B}= ( U^i/U^t)_{_B}$. In the Newtonian limit, Eqs.~(\ref{coeqns1}-\ref{etensor}) lead to the Euler equation (see Ref.~\cite{Weinberg1972})
\begin{align}
\frac{\partial {\bf v}_{_B}}{\partial t} + ({\bf v}_{_B}\cdot \nabla){\bf v}_{_B} &= -\frac{1-{\bf v}^2_{_B}}{\bar\rho_{_B} + \bar p_{_B}}\left[ \nabla \bar p_{_B}+ {\bf v}_{_B} \frac{\partial \bar p_{_B}}{\partial t}\right]\label{euler}\\
& + {\rm terms\,\,of\,\,long\!\!-\!\!range \,\,forces}.
\nonumber
\end{align}
The first term in the right-handed side of Eq.~(\ref{euler}) indicates the force due to the space-time gradients of baryon fluid pressure. This implies that the space-time gradients of baryon fluid velocity ${\bf v}_{_B}({\bf x},t)$ should have the rates of short-range strong interactions, which are proportional to the inverses of $\pi$, $\sigma$, $\rho$ and $\omega$ meson masses ($\sim m^{-1}_{\pi,\sigma,\rho,\omega,\cdot\cdot\cdot}$), depending on values of the baryon density $\bar n_{_B}({\bf x},t)$. These nuclear reaction rates must be larger than the rate ($\gtrsim m_e^{-1}$) of electromagnetic interactions. In other words, the baryon fluid and electron fluid have the different values of the incompressibility so that they have different rates (frequencies) of reactions in space and time. However, this still remains as an argument, because we has not so far been able to quantitatively calculate the space-time gradients of baryon fluid pressure by Eqs.~(\ref{p_b}) and (\ref{e_b}), then to obtain the space-time gradients of baryon fluid velocity by Euler equation (\ref{euler}) together with the Einstein-Maxwell field equations. 
\comment{Nevertheless, in Ref.~\cite{hrx2012} using this argument as an assumption for the baryon fluid velocity ${\bf v}_{_B}({\bf x},t)$ at different selected values of collapsing radii, we numerically solve equations for the number and energy-momentum conservations of electron fluid and Maxwell equations coupling to the electric current of baryon fluid. As a result, we show the oscillating phenomenon of electron fluid and electric fields leading to electron-positron pair production, and calculate the rates of these electromagnetic processes. 
}

In the following, we attempt to address our attention to the issue how the gravitational energy gained by the baryon fluid in collapses is transfered to the electromagnetic energy and how kinetic and internal energies are reduced as a consequence of total energy conservation.   
The energy conservation (\ref{coeqns1}) along a flow line of the electron fluid yields
\begin{align}
U^\mu_e(T^{\nu}_{\,\,\,\mu})_{;\nu} =e\bar n_p F_{\mu\nu}U^\mu_eU^{\nu}_B =e\bar n_p\gamma_e\gamma_B(v_B-v_e)g_{rr}E,
\label{decay}
\end{align}
where $e$ and $E$ are electric charge and field, the fluid velocity $v_{(e,B)}\equiv v^r_{(e,B)}= ( U^r/U^t)_{(e,B)}$ and Lorentz factor $\gamma_{(e,B)}\equiv (1+U_rU^r)^{1/2}_{(e,B)}$ in the spherical geometry 
\begin{align}
&&ds^2 =-g_{tt}dt^2+g_{rr}dr^2+r^2d\theta^2 +r^2\sin^2\theta
d\phi^2 ~. \label{sw}
\end{align}
Eq.~(\ref{decay}) indicates that the dynamical evolutions of the baryon fluid caused by the gravitational or strong interactions can transfer the energy that the baryon fluid gains to the
electron fluid via an electric field, provided $v_e\not\equiv v_B$. 
As explained in the introductory section, for the reason that the differential equations governing macroscopic processes (e.g.~gravitational collapse) and the differential equations governing microscopic processes 
(e.g.~electrodynamic pair-production, nuclear reaction) have very different space-time scales at least of the order of $10^{17}$, it is very difficult to simultaneously integrate these differential equations and quantitatively show the energy transformation as indicated by Eq.~(\ref{decay}) in the realistic case of gravitational collapses. In order to overcome these difficulties and make steps toward the understanding of the issue, on the basis of some assumptions and approximations, we decouple the differential equations governing macroscopic processes from the differential equations governing microscopic processes as follows.

\begin{enumerate}

\item The first, we study the static case of compact stars at/over the nuclear density, e.g., baryons and electrons of neutral compact stars are in their equilibrium states. The local equilibrium profile of baryons must be determined by the strong interaction, whereas the local equilibrium profile of electrons must be determined by the electromagnetic interaction. In the Thomas-Fermi model, an overcritical ``equilibrium'' electric fields are found \cite{Usov1} on the surface of baryon cores. These results provide the initial configurations for the dynamical space-time evolution of baryon core and electron fluid in the gravitational collapse or pulsation.

\item Because of the dynamics of gravitational collapse or pulsation, the baryon core deviates from its equilibrium state. We postulate that due to the nuclear rigidity of baryon cores, an inward velocity $v_B$ and charged current $J_B$ (Eqs.~(9) in \cite{hrx2012}) of baryon cores are introduced at the rate of the nuclear reaction scales, rather than the rate of the gravitational collapse, as already indicated in Eqs.~(\ref{p_b},\ref{e_b},\ref{euler}).
We asked the question how the electron fluid responses to this external baryon current $J_B$. In Ref.~\cite{hrx2012}, by solving the microscopic kinetic transport equations (particle number and energy-momentum conservation) of the electron fluid as well as the Maxwell equations, we obtained the space-time evolution (non-equilibrium) of the electron fluid and overcritical electric fields in the Compton scale, and estimated the rate of pair-productions. These results are essentially due to the postulation that the inward baryon current $J_B$ is introduced at the rate of the strong interaction scale, rather than the gravitational one. The rate of gravitational collapses is too slow to trigger these electrodynamic processes at the Compton scale. In addition, it should be pointed out that in these calculations we did not solve the differential equations for the electron fluid and the Maxwell equation together with the differential equation for the gravitational collapse. The baryon velocity $v_B$ is treated as a parameter and its values are given by a simple collapsing equation of thin shell at different radii of gravitational collapse (Figure 3 in Ref.~\cite{hrx2012}). In summary, two important assumptions were made: (i)  the baryon core is treated as a giant nucleus and the deviation from its equilibrium state, represented by the baryon electric current $J_B\sim v_B$, is introduced at the rate of the strong introduction; (ii) the values of $v_B$ are given by a simple collapsing model without considering dynamics of the gravitational collapses. 

\item On the contrary, instead of solving the differential equations for microscopic electrical processes in a given dynamics of gravitational collapse, in this article we focus on solving differential equations for macroscopic gravitational collapse processes in a given dynamics of electric processes studied in Ref.~\cite{hrx2012}, represented by an ansatz function. Our purpose is to see the back-reaction of microscopic electrical processes on macroscopic gravitational collapse processes. In order to gain some insight into this issue, we study the gravitational collapse of a spherical thin-shell capacitor, which might present a thin layer of collapsing stellar cores. Although this spherical thin-shell capacitor is totally neutral, it carries electric and gravitational energies. Using such a simplified model, we try to find an analytical description and make a step in understanding the issue how the gravitational energy is converted to electric, kinetic and internal energies in a neutral stellar core collapse.

\end{enumerate}

This has been so far our approach to the electromagnetic processes in the gravitational collapse of neutral compact stars at/over nuclear density. This approach is clearly far from being complete. In order to quantitatively show that the production, oscillation and annihilation of electron-positron pairs with overcritical electric fields indeed dynamically take place, one must solve altogether the Maxwell equation and the quantum Boltzmann-Vlasov transport equations not only for the electrons fluid \cite{RVX03a}, but also for the baryon fluid with the strong interaction. We have not yet been able to model the strong interaction for doing these quantitative calculations. On the basis of the rates of various microscopic processes and interactions, we argue the possibility of the production, oscillation and annihilation of electron-positron pairs and dynamical evolution of overcritical electric fields (Eq.~(26) in Ref.~\cite{hrx2012}). We clarify that in our model these electric processes are triggered by the rapid action rate of the baryon core due to the strong interaction, rather than the gravitational interaction. However, the question is to understand how to quantitatively describe and calculate the dynamics of strongly interacting baryon core in gravitational collapse, and how the baryon charged current $J_B$ is introduced at the rate of strong interactions. This will be the subject for our future work.      

\noindent
{\bf A spherical thin-shell capacitor}
\hskip0.1cm   
The spherical thin-shell capacitor is composed by a layer of positively charged baryons and a layer of negatively charged electrons. The baryon layer is defined as a mathematically thin layer, while the electron layer is understood as a physically thin layer with a thickness ``$d$'' specified below. The total numbers of charged baryons and electrons are exactly equal so that the spherical thin-shell capacitor is totally neutral but carries non-vanishing the electric energy stored inside two spherical layers.
The number-densities of two spherical layers are at least order of the nuclear density, as a consequence the radial separation ``$d$'' between two spherical layers must be a few orders of the Compton length $\lambda_C$. The reasons are the following: electric fields between two layers $E\approx e \bar n_{\rm nucl} d$ are overcritical and electric force acting on ultra-relativistic electrons balances their Fermi momenta $eEd\approx P_e^F\approx \bar n_{\rm nucl}^{1/3}$. Let the baryon layer locate at the Schwarzschild-like radial coordinate $r_{0}$ and electron layer distributes from $r_0$ to $r_0+d$. The spherical thin-shell capacitor can be physically considered as an infinitely thin shell for $d/r_0 \rightarrow 0$. The spherical 
thin-shell capacitor is henceforth denoted by ``the thin shell'' in short.
 
As the baryon layer is mathematically thin, in Eq.~(\ref{etensor}) the baryon pressure $\bar p_B=0$ and mass density 
$\bar \rho_B(x)=\bar \rho_B\delta^{\left(  4\right)  }\left(  x,x_0\right)$, where $\bar \rho_B$ is the constant surface density in the proper frame of the baryon layer and the $4$-dimensional Dirac distribution is defined as
\begin{equation}
\int\delta^{\left(  4\right)  }\left(  x,x_0\right)  \sqrt{-g}d^{4}x=1,
\nonumber
\end{equation}
where $g=\det\left\Vert g_{\mu\nu}\right\Vert $.
\comment{ 
and $\delta^{\left(  4\right)  }\left(  x,x_0\right)=(1/g)\delta (t-t_0)\delta(r-r_0)\delta(\theta-\theta_0)\delta(\phi-\phi_0)=(1/r^2(-g_{tt}g_{rr})^{1/2})\delta (t-t_0)\delta(r-r_0)\delta(\cos\theta-\cos\theta_0)\delta(\phi-\phi_0)$. 
The energy-momentum tensor of spherical thin-shell capacitor $T^{\mu\nu}_{(e,B)} = (\bar\rho_{(e,B)})U^\mu_{(e,B)} U^\nu_{(e,B)}$, the pressure $\bar p_{(e,B)}=0$ in Eq.~(\ref{etensor}).
} Then we have ($d\Omega=\sin\theta d\theta d\phi$)
\begin{align}
\int\bar\rho_B\delta^{\left(  4\right)  }\left(
x,x_{0}\right)  r^{2}drd\Omega d\tau &= M_{0},\label{eq1a}
\end{align}
where $M_{0}$ is the rest mass of the baryon layer,
and $\tau$ is the proper time along the world surface $S:$ $x_{0}=x_{0}\left(
\tau,\theta, \phi \right)$ of the baryon layer. $S$ divides the space-time into two
complementary static space-times: an internal one $\mathcal{M}_{-}$ and an external one $\mathcal{M}_{+}$. Their time-like Killing vectors are denoted by $\xi^\mu_-$ and $\xi^\mu_+$. $\mathcal{M}_{+}$ is foliated by the family $\{\Sigma_t^+: t_+=t\}$ of space-like hypersurfaces of constant $t_+$. 

On the other hand, introducing the orthonormal tetrad
\begin{equation}
{\boldsymbol{\omega}}_{\pm}^{\left(  0\right)  }=(g_{tt}^{\pm})^{1/2}dt,\quad
{\boldsymbol{\omega}}_{\pm}^{\left(  1\right)  }=(g_{rr}^{\pm})^{-1/2}dr,\quad
{\boldsymbol{\omega}}^{\left(  2\right)  }=rd\theta,\quad{\boldsymbol{\omega}%
}^{\left(  3\right)  }=r\sin\theta d\phi,
\end{equation}
we describe the electric field ${\boldsymbol{E}}=E{\boldsymbol{\omega}}^{(1)}$ and electromagnetic tensor $(T^{\mathrm{em}}){}{}_{t}{}^{t}= E^2/(8\pi)$ and $(T^{\mathrm{em}}){}{}_{i}{}^{i}= -E^2/(8\pi)$ inside the thin shell ($r_0\le r\le r_0+d$). 
\comment{
has strength%
\begin{equation}
E^{\,\left(  1\right)  }_{\pm}=Q_{\rm eff}\,\frac{\delta\left(r-r_0\right)}{g}=\frac{Q_{\rm eff}}{r^2}\delta\left(r-r_0\right)\,,
\label{E3}%
\end{equation}
where $\delta\left(r-r_0\right)/g$ is the spherically symmetric $\delta$-function in the space-like hypersurfaces $\{\Sigma_t^\pm: t_\pm=t\}$,
}  
The
electric energy of the thin shell, measured by an observer at rest at infinity, is obtained by evaluating the Killing
integral
\begin{equation}
\int_{\Sigma_{t}^{+}}\xi_{+}^{\mu}T_{\mu\nu}^{\mathrm{em}}d\Sigma^{\nu}_{+}%
=4\pi \int_{r}^{\infty}r^{2}dr\ (T^{\mathrm{em}}){}{}_{t}{}^{t}\equiv \frac{Q^2_{\rm eff}(r)}{2r}, \label{EQR}%
\end{equation}
where $d\Sigma^{\nu}_{+}$ is the
surface element vector of the space-like hypersurfaces $\Sigma_{t}^{+}$ in $\mathcal{M}_{+}$. In Eq.~(\ref{EQR}), we introduce the quantity $Q^2_{\rm eff}(r)\not= 0$ for $ r_0\leq r\leq r_0+d$ to characterize the electric energy stored inside the thin shell.  $Q^2_{\rm eff}(r)=0$ for $r > r_0+d$ and $r< r_0$. The total electric energy inside the thin shell is given by
\begin{equation}
{\mathcal E}_{\rm em}(r_0)=\frac{Q^2_{\rm eff}(r_0)}{2r_0},
\label{e-energy}
\end{equation} 
where the quantity $Q^2_{\rm eff}(r_0)$ parametrizes the total electric energy stored inside the thin shell that locates at radius $r_0(t_0)$ and time $t_0$. $Q_{\rm eff}(r)$ does not represent an electric charge carried by the thin shell. We express the repulsive electric energy (\ref{EQR}) or (\ref{e-energy}) in the same form of the Coulomb energy of a spherical charged layer for the reason that it is useful to study the collapse equation of the thin shell in next section. It is worthwhile to mention again that if we do not consider the strong interaction that differentiates between protons and electrons, electric charges must be complete screened so that the electric field and energy vanish and $Q^2_{\rm eff}$ of Eq.~(\ref{EQR}) is zero. In this case, we are led to the traditional scenario describing the collapse of neutral thin shell.       

The energy-momentum tensor (\ref{etensor_e}) of the electron layer has a physical distribution over the size ``$d$'' of the thin shell. Analogously to Eq.~(\ref{EQR}), we define the total energy of the electron layer as 
\begin{equation}
{\mathcal E}_{\rm electron}(r_0)\equiv \int_{\Sigma_{t}^{+}}\xi_{+}^{\mu}(T_e)_{\mu\nu}d\Sigma^{\nu}_{+}%
=4\pi \int_{r_0}^{\infty}r^{2}dr\ (T_e){}{}_{t}{}^{t}, 
\label{EQR_e}%
\end{equation}
where $(T_e)^{t\,t}=(\bar\rho_e + \bar p_e \langle {\bf v}^2_e\rangle )/(1-\langle {\bf v}^2_e\rangle)$ and ${\bf v}_e$ is the electron fluid velocity. In Ref.~\cite{hrx2012}, it is shown that the electron fluid velocity ${\bf v}_e$ is ultra-relativistically oscillating back and forth collectively with oscillating electric fields inside the thin shell, $\langle {\bf v}^2_e\rangle$ indicates the averaged value over rapid oscillations in the Compton scale. In Eq.~(\ref{EQR_e}), the rest mass of the electron layer is negligible, compared with its internal energy for ultra-relativistically oscillating electrons. Moreover, at or over the nuclear density, electron Fermi momenta $P^F_e\sim m_\pi $ in the proper frame of the electron fluid is rather smaller than the baryon mass $m_{_B}$. Therefore, compared with the rest mass of baryon layer $M_0$, we neglect the internal energy of electron layer ${\mathcal E}_{\rm electron}(r_0)$ of Eq.~(\ref{EQR_e}) in this article.

The thin baryon shell collapses in the velocity $dr_0/dt_0$ which will be discussed in the next section.  The electron fluid collapses together with the baryon shell, due to strong Coulomb force, however the electron fluid velocity ${\bf v}_e$ has two components: (i) the baryon shell velocity $dr_0/dt_0$; (ii) the oscillating velocity  inside the baryon shell. 
In the approximation of thin shell model, we disregard the detailed space-time oscillations of electron fluid and  electric field in the Compton length scale, which lead to the energy radiation in the form of electron-positron pairs. Instead, we attempt to properly model the quantity $Q^2_{\rm eff}(r_0)$ to represent these microscopic processes of building the electric energy (\ref{e-energy}) and radiating it away from the thin shell, so as to study the back-reaction of these microscopic processes on the macroscopic process of gravitational collapse of the thin shell.

\comment{
Actually the total electric energy (\ref{EQR}) can be described by the prescription of $\delta$-function for the electric energy density, analogously to Eq.~(\ref{eq1a}).    
As will be discussed below, the net charge $Q_{\rm net}(r)$ and electric field $E(r)$ are varying functional in collapses, and equilibrium configurations $Q^{\rm eq}_{\rm net}(r)$ and $E_{\rm eq}(r)$ are for the static case. As a result, the total electric energy ${\mathcal E}_{\rm em}=Q^2_{\rm eff}(r_0)/2r_0$ is varying functional in collapses.   
}

\noindent
{\bf Collapse of spherical thin-shell capacitor}
\hskip0.1cm 
A lot of attention has been focused on the exact solution of thin charged shell in gravitational collapse \cite{I66}. Following the line presented in Refs.~ \cite{crv2002} and \cite{RV2002} for finding an exact solution of thin charged shell in gravitational collapse, we try to approximately solve the Einstein equations (\ref{coeqns1},\ref{tem}) for the gravitational collapse of the spherical thin-shell capacitor (the thin shell). We have $g_{tt}^{-}= (g_{rr}^{-})^{-1}\equiv f_{-}$ and $g_{tt}^{+} \approx (g_{rr}^{+})^{-1}\equiv f_{+}$, where the sign ``$\approx$'' indicates for the range $r_0\geq r\geq r_0+d$, where we neglect the charge and mass-energy distributions of the electron layer.  From the $G_{tt}$
Einstein equation, we get
\begin{equation}
ds^{2}=\left\{
\begin{array}
[c]{l}%
-f_{+}dt_{+}^{2}+f_{+}^{-1}dr^{2}+r^2(d\theta^2 +\sin^2\theta
d\phi^2)\qquad\text{in $\mathcal{M}_{+}$}\\
-f_{-}dt_{-}^{2}+f_{-}^{-1}dr^{2}+r^2(d\theta^2 +\sin^2\theta
d\phi^2)\qquad\text{in $\mathcal{M}_{-}$}%
\end{array}
\right.  , \label{E0}%
\end{equation}
where 
\begin{equation}
f_{+}=1-\tfrac{2M}{r}+\tfrac{Q^2_{\rm eff}(r)}{r^2},\quad {\rm and} \quad f_{-}=1;
\label{f+-}
\end{equation}
$t_{-}$ and $t_{+}$ are the
Schwarzschild-like time coordinates in $\mathcal{M}_{-}$ and $\mathcal{M}_{+}$
respectively. $M$ is the total mass-energy of the thin shell, measured by an observer at rest at infinity.
Indicating by $t_{0\pm}$ the Schwarzschild-like time coordinate of the thin shell, from the $G_{tr}$ Einstein equation we
have 
\begin{equation}
\tfrac{M_{0}}{2}\left[  f_{+}\left(  r_{0}\right)  \tfrac{dt_{0+}}{d\tau
}+f_{-}\left(  r_{0}\right)  \tfrac{dt_{0-}}{d\tau}\right]  =M%
-\tfrac{Q_{\rm eff}^{2}}{2r_{0}}, \label{eq3a}%
\end{equation}
where we introduce the notation $Q_{\rm eff}^{2}\equiv Q_{\rm eff}^{2}(r_0)$.
The remaining Einstein equations are identically satisfied. From (\ref{eq3a}) we have that the inequality
\begin{equation}
M-\tfrac{Q_{\rm eff}^{2}}{2r_{0}}>0, 
\label{Constraint}%
\end{equation}
holds since the left-handed side of Eq.~(\ref{eq3a}) is clearly positive.
We define the four-velocity $U^\mu$ of the thin shell as the four-velocity $U^\mu_B$ of the baryon layer, for the reasons discussed in the paragraphs where Eqs.~(\ref{p_b}-\ref{euler}) are. From (\ref{eq3a})
and the normalization condition of the four-velocity of the thin shell $U_{\mu}U^{\mu}=-1$,
\begin{equation}
\left[  -f_{\pm}\left(  r_{0}\right)  \tfrac{dt_{0\pm}}{d\tau
}+f_{\pm}\left(  r_{0}\right)  \tfrac{dt_{0\pm}}{d\tau}\right] =-1, \label{U-con}%
\end{equation}
we find
\begin{align}
\left(  \tfrac{dr_{0}}{d\tau}\right)  ^{2}  &  =\tfrac{1}{M_{0}^{2}}\left(
M\pm\tfrac{M_{0}^{2}}{2r_{0}}-\tfrac{Q_{\rm eff}^{2}}{2r_{0}}\right)  ^{2}-f_{\mp}\left(  r_{0}\right),\label{EQUY}\\
\tfrac{dt_{0\pm}}{d\tau}  &  =\tfrac{1}{M_{0}f_{\pm}\left(  r_{0}\right)
}\left(  M\mp\tfrac{M_{0}^{2}}{2r_{0}}-\tfrac{Q_{\rm eff}^{2}}{2r_{0}}%
\right)  , \label{EQUYa}%
\end{align}
in the space-times $\mathcal{M}_{\pm}$. Eqs.~(\ref{E0}-\ref{EQUYa}) 
completely describe a 3-parameter ($M$, $Q^2_{\rm eff}$, $M_{0}$) family
of solutions of the Einstein equations.
\comment{ 
The equation of motion of the shell (\ref{EQUY}) can also be written as 
\begin{equation}
M=M_0\sqrt{1+\left(\frac{dr_0}{d\tau}\right)^2}  
+\tfrac{Q_{\rm eff}^{2}}{2r_{0}}\mp \tfrac{M_0^{2}}{2r_{0}}, \label{eq3a}%
\end{equation}
in the space-times $\mathcal{M}_{\pm}$. 
The equation of
motion for the shell, Eq.~\ref{EQUY}, reduces in this case to
\begin{equation}
\left(  M_{0}\tfrac{dr_{0}}{d\tau}\right)  ^{2}=\left(  M+\tfrac{M_{0}^{2}%
}{2r_{0}}-\tfrac{Q^2_{\rm eff}}{2r_{0}}\right)  ^{2}-M_{0}^{2} \label{EQA}%
\end{equation}
in the space-times $\mathcal{M}_{-}$ and
\begin{equation}
\left(  M_{0}\tfrac{dr_{0}}{d\tau}\right)  ^{2}=\left(  M-\tfrac{M_{0}^{2}%
}{2r_{0}}-\tfrac{Q^2_{\rm eff}}{2r_{0}}\right)  ^{2}-M_{0}^{2}f_{+}(r_0) \label{EQAb}%
\end{equation}
in $\mathcal{M}_{+}$.
} 
As we will see, for the description of the collapse
we can choose either $\mathcal{M}_{-}$ or $\mathcal{M}_{+}$. The two
descriptions are equivalent and relevant for the physical
interpretation of the solutions.

For astrophysical applications, see for example Ref.~\cite{RVX03c}, we attempt to approximately solve the equation of motion of the thin shell and obtain the trajectory 
$r_{0}=r_{0}\left(  t_{0+}\right)$ as a function of the time
coordinate $t_{0+}$ relative to the space-time region $\mathcal{M}_{+}$. In
the following we drop the $+$ index from $t_{0+}$. From (\ref{EQUY}) and
(\ref{EQUYa}) we have the equation of motion of the thin shell
\begin{align}
\tfrac{dr_{0}}{dt_{0}}&=\tfrac{dr_{0}}{d\tau}\tfrac{d\tau}{dt_{0}}=\pm\tfrac
{F}{\Omega}\sqrt{\Omega^{2}-F},\nonumber\\ \tfrac{dr_{0}}{d\tau}&=\pm\sqrt{\Omega^{2}-F} \label{EQUAISRDLC}%
\end{align}
where $F\equiv f_{+}\left(  r_{0}\right)$ of Eq.~(\ref{f+-}),
\begin{equation}
\Omega\equiv\Gamma-\tfrac{M_{0}^{2}+Q_{\rm   eff}^{2}}%
{2M_{0}r_{0}},\quad\Gamma\equiv\tfrac{M}{M_{0}}.
\end{equation}
Since we are interested in an imploding thin shell, only the minus sign case in
(\ref{EQUAISRDLC}) will be studied. We can give the following physical
interpretation of $\Gamma$. For $M\geq M_{0}$, $\Gamma$ coincides with
the Lorentz factor of the imploding thin shell at infinity; from
(\ref{EQUAISRDLC}) it satisfies
\begin{equation}
\Gamma=\tfrac{1}{\sqrt{1-\left(  \frac{dr_{0}}{dt_{0}}\right)  _{r_{0}=\infty
}^{2}}}\geq1.
\end{equation}
We rewrite equation of motion (\ref{EQUAISRDLC}) as
\begin{align}
\left(\frac{dr_0}{d\tau}\right)^2=\left[\Gamma + \frac{M_0}{2r_0}(1-\xi^2)\right]^2-1,
\nonumber
\end{align}
or
\begin{align}
\left(\frac{\Omega}{F}\right)^2\left(\frac{dr_0}{dt_0}\right)^2=\left[\Gamma + \frac{M_0}{2r_0}(1-\xi^2)\right]^2-1,
\label{coll}
\end{align}
where $\Omega\equiv \Gamma-(M_0/2r_0)(1+\xi^2)$ and we define an effective ``charge-mass-ratio'' 
\begin{align}
\xi \equiv \frac{Q_{\rm eff}}{M_0}.
\label{effx}
\end{align}
Actually $\xi^2$ represents the ratio of electric energy and gravitational energy of the thin shell.
For the case $\Gamma=1$ ($M=M_0$), i.e., the thin shell collapses at rest from infinity. Eq.~(\ref{Constraint}) requires $M_0\ge Q_{\rm eff}^{2}/2r_0$ to start gravitational collapse and Eq.~(\ref{coll}) requires $\xi < 1$ to continue gravitational collapse. When $\xi=1$, gravitational collapse stops and kinetic energy of the thin shell vanishes as will be seen below. The trajectory of the thin shell is given by the solution:
\begin{equation}
\int dt_{0}=-\int\tfrac{\Omega}{F\sqrt{\Omega^{2}-F}}dr_{0}. \label{GRYD}%
\end{equation}
to the equation of motion (\ref{EQUAISRDLC}).

To understand the total energy conservation of the thin shell in gravitational collapse, we use the solution (\ref{EQUY}) in the flat space-time  $\mathcal{M}_{-}$,
\begin{align}
\left(  M_{0}\tfrac{dr_{0}}{d\tau}\right)  ^{2}  &  =\left(
M+\tfrac{M_{0}^{2}}{2r_{0}}-\tfrac{Q_{\rm eff}^{2}}{2r_{0}}\right)  ^{2}-M_{0}^2,\label{EQUY-}
\end{align}
we can interpret $-\tfrac
{M_{0}^{2}}{2r_{0}}$ as the gravitational attractive energy of the thin
shell and $\tfrac{Q^{2}_{\rm eff}}{2r_{0}}$ is its repulsive electric energy. 
Introducing the total four-momentum of the shell $P^\mu=M_0U^\mu$ and its radial component $P\equiv M_{0}U^{r}=M_{0}\tfrac{dr_{0}%
}{d\tau}$, the kinetic energy of the thin shell as
measured by static observers in $\mathcal{M}_{-}$ is expressed as \cite{RV2002}
\begin{align}
T(r_0)\equiv -P_{\mu}%
\xi_{-}^{\mu}-M_{0}=\sqrt{P^{2}+M_{0}^{2}}-M_{0}. 
\label{EQUY-k}
\end{align}
Then from Eqs.~(\ref{EQUY-},\ref{EQUY-k}) we have
\begin{align}
M(r_0)&= -\tfrac{M_{0}^{2}}{2r_{0}}+\tfrac{Q^{2}_{\rm eff}}{2r_{0}}+\sqrt{P^{2}+M_{0}^{2}}\nonumber\\
&=M_{0}+T(r_0)-\tfrac{M_{0}^{2}}{2r_{0}}+\tfrac{Q^{2}_{\rm eff}}{2r_{0}}, \label{EQC}%
\end{align}
where we choose the positive root solution due to the constraint (\ref{Constraint}).
Eq.~(\ref{EQC}) is the total energy-conservation of the thin shell, whose rest mass $M_0$, kinetic energy $T(r_0)$, gravitational energy $-\tfrac{M_{0}^{2}}{2r_{0}}$, and electric energy $\tfrac{Q^{2}_{\rm eff}}{2r_{0}}$ depend on the radial coordinate $r_{0}(t_0)$ in gravitational collapse. 

In the following discussion, we consider the shell is at rest at infinity and starts to gravitational collapse, $T(r_0)=0$, $-\tfrac{M_{0}^{2}}{2r_{0}}=0$ and $\tfrac{Q^{2}_{\rm eff}}{2r_{0}}=0$ at $r_0\rightarrow \infty$. The initial energy of the thin shell $M(r_0\rightarrow \infty)=M_0$, i.e., $\Gamma=1$. The total shell energy $M(r_0)=M_0$ is conserved in the entire collapsing process. 
\comment{
Eq.~(\ref{EQC}) becomes in unit of $M_0$
\begin{align}
M(r_0)
&=M_{0}+T(r_0)-\tfrac{M_{0}^{2}}{2r_{0}}+\tfrac{Q^{2}_{\rm eff}}{2r_{0}}, \label{EQC1}%
\end{align}
which explicitly shows that the gravitational energy gained by the collapsing shell becomes the kinetic energy and electric energy of the shell.
}

\noindent
{\bf Collapse of the thin shell with varying electric energy}
\hskip0.1cm
In Ref.~\cite{hrx2012}, assuming that in gravitational collapses, the baryon layer induces an inward current-density 
\begin{align}
J_{_B}^r(r_0)=e\bar n_pU^r_{_B}\approx e\bar n_p(\dot r_0\Omega/F),\quad \dot r_0 =dr_0/dt_0,
\label{cur}
\end{align}
at the rate of strong interaction scales, we show that triggered by this baryon current (\ref{cur}),
the current-density $J_e^r=e\bar n_eU^r_e$ of the electron layer oscillates collectively with overcritical electric fields $E$ at frequency $\omega_{\rm osci} = \tau^{-1}_{\rm osci}\simeq 1.5\, m_e$, leading to the production of electron-positron pairs at rate $\tau^{-1}_{\rm pair}\simeq 6.6\, m_e$. Selecting values $J_{_B}^r(r_0)$ and $\dot r_0$ of Eq.~(\ref{cur}) at different collapsing radii, we calculated \cite{hrx2012} the averaged energy and number densities of electron-positron pairs produced, as well as the averaged electric energy (Coulomb energy) of oscillating overcritical electric fields. In addition, our results presented in Refs.~\cite{RVX03a,RVX03c} show that these electron-positron pairs annihilate to photons and the ultra-dense plasma of electron-positron pairs and photons is formed with the equipartition of energy and number of electron-positron pairs and photons, beside this plasma undergoes the hydrodynamical expansion and the photon radiation occurs. This indicates that the electric energy is established by the electron-positron oscillations collectively with overcritical electric fields, then dissipated by electron-positron annihilations to photons radiating away. Clearly, these results and discussions are based on the postulation that the baryon current of Eq.~(\ref{cur}) introduced by the strong interaction in a gravitational collapse process triggers all electric processes, provided that the reaction rates of processes satisfy the inequality of Eq.~(26) in Ref.~\cite{hrx2012}. In the light of the total energy conservation in gravitational collapses and Eq.~(\ref{decay}), we further postulate that the electric energy of these electric processes is converted from the gravitational energy, as a consequence, the gravitational energy gained by the collapsing baryon core is transfered to the photon radiation energy. In future work, we are bound to show this energy conversion by solving the equations of gravitational collapses altogether with the equations of electric processes and nuclear processes. In the present article, we attempt to study the back-reaction effect of this energy conversion on the gravitational collapse. 
  
In the simplified model of collapsing thin shell, we represent $\tfrac{Q^{2}_{\rm eff}}{2r_{0}}$ the electric energy established by electron-positron pair production and oscillation with overcritical electric fields, then dissipated by electron-positron annihilations to photons radiating away at the collapsing radius $r_0$. The time variation rate of this electric energy $\tfrac{Q^{2}_{\rm eff}}{2r_{0}}$ is characterized by the frequency $\omega_{\rm osci} \simeq 1.5 m_e$ \cite{hrx2012}. On the other hand,
from collapse equation (\ref{coll})  for $\Gamma=1$, it is shown that the collapsing velocity $(dr_0/dt_0)$ varies between zero and its maximal value as the ``charge-mass-ratio'' $\xi$ varies from $1$ and $0$, corresponding to the microscopic processes of the electric energy $\tfrac{Q^{2}_{\rm eff}}{2r_{0}}$ built up and completely radiating away. On the basis of numerical results (Fig.~5) in Ref.~\cite{hrx2012},
we model the varying ``charge-mass-ratio''
\begin{align}
\xi = \xi^{\rm max} |\sin (\omega_{\rm osci} r_0)| + \xi^{\rm min},\quad r_0=r_0(t_0). 
\label{xdecay}
\end{align}
where $\omega_{\rm osci} \simeq 1.5 m_e$, $\xi^{\rm max}=0.6$ and $\xi^{\rm min}=0.1$, to illustrate the back-reaction of this radiative electric energy on the gravitational collapse of the thin shell. This postulates that at the collapsing radius $r_0(t_0)$ of the baryon layer, the microscopic processes of the electric energy $\tfrac{Q^{2}_{\rm eff}}{2r_{0}}$ built up and radiating away are in the rate of the Compton scale $\omega_{\rm osci} \simeq 1.5 m_e$ and effectively described by a simple function of Eq.~(\ref{xdecay}), and $\xi^{\rm min}\not= 0$ representing 
the part of the electric energy that does not radiate away from the shin shell. Whereas the case ($\xi\equiv 0$) represents the collapse of a neutral thin shell without carrying any electric energy. 

\begin{figure}[th]
\begin{center}
\includegraphics[width=8cm]{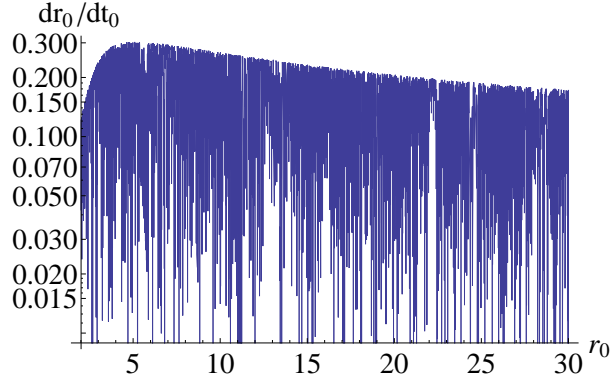}
\end{center}
\caption{In unit of the speed of light $c$, the collapse velocity $(dr_0/dt_0)$ is plotted (fast oscillating lines in blue) as a function of radius $r_0$ of the collapsing thin shell. The thin shell is at rest at the radius $R_0=30\, GM_0$ and starts to collapse. The thin shell mass  $M_0=20M_{\odot}$.}%
\label{gc_fig1}%
\end{figure}

We express $r_0$ and $t_0$
in units of $GM_0$ and $GM_0/c$, then $\omega r_0 = 1.5 (m_e GM_0)r_0$, $\lambda_C /GM_0= 1.05\times 10^{-16}$, $20 GM_\odot /c^2\simeq 10^{-4}$ second and $M_0=20 GM_\odot/c \simeq 3\times 10^{6}\,$cm.
Plotting the velocity $\dot r_0 = dr_0/dt_0$ of Eq.~(\ref{coll}) in Fig.~\ref{gc_fig1}, we find that in collapse process, the thin shell velocity is oscillating between zero and the envelop curve, which represents the collapsing velocity of the thin shell carrying the electric energy described by $\xi^{\rm min}\not= 0.1$. This result shows a sequence of ``on and off'' collapsing steps: the thin shell at rest starts to move inwards due to the gravitational attraction of the baryon layer, and stops due to the repulsion of the electric energy $\tfrac{Q^{2}_{\rm eff}}{2r_{0}}$ built up to $\xi=1$, then restarts to move inwards again due to the electric energy $\tfrac{Q^{2}_{\rm eff}}{2r_{0}}$ partially radiating away in the form of electron-positron pairs and photons. The frequency of this ``on and off'' hopping sequence is about $\omega_{\rm osci} \sim  m_e$, the Compton scale. The collapse process is still continuous in terms of macroscopic scale. However,
as will be seen soon, the time scale and kinetic energy of collapses are changed. 

\begin{figure}[th]
\begin{center}
\includegraphics[width=8cm]{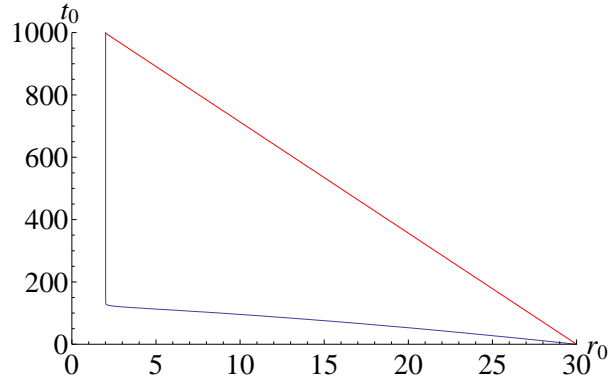}
\end{center}
\caption{In thin shell collapsing process, the time coordinate $t_0$ is plotted as a function of radial coordinate $r_0$ of the thin shell. $t_0$ and $r_0$ are in units of $GM_0$. The red line is for $\xi$ of Eq.~(\ref{xdecay}) and the blue for $\xi=0$. The shell is at rest at the radius $R_0=30\, GM_0$ and starts to collapse. The thin shell mass  $M_0=20M_{\odot}$.} %
\label{gc_fig2}%
\end{figure}

The averaged collapsing velocity of the thin shell of Eq.~(\ref{xdecay}) is smaller than the collapsing velocity 
(envelop curve) for the case $\xi = 0$. As a result, the time duration of collapse process becomes longer.  Assuming that the thin shell is at rest at the radius $R_0=30M_0$ and starts to collapse, we plot in Fig.~\ref{gc_fig2} the time coordinate $t_0$ of Eq.~(\ref{GRYD}) as a function of the radial coordinate $r_0$ of the collapsing thin shell, in comparison with that of the case $\xi = 0$. The blue line for the case $\xi=0$ shows that the collapsing shell takes time $\sim 10^2\, GM_0/c^2$  to approach the horizon, whereas the red line for the case $\xi$ of Eq.~(\ref{xdecay}) shows that the collapsing thin shell takes time $\sim 10^3\, GM_0/c^2$ to approach the horizon. The collapsing time for the case $\xi$ of Eq.~(\ref{xdecay}) is about $10$ times longer than the collapsing time for the case $\xi = 0$. This result is not sensitive to the value of the frequency $\omega_{\rm osci}$ in the Compton scale and the detailed form of an oscillating function (\ref{xdecay}) of the frequency $\omega_{\rm osci}$. 

It should be pointed out that in this simplified toy model of thin shell collapsing, to evidently illustrate the back-reaction effect that slows down the collapsing process in comparison with the free fall collapsing process in the same plot (see Fig.~\ref{gc_fig2}), we select the initial radius $R_0=30M_0$ at which the thin shell starts to collapse. As discussed, the baryon core must be at (over) the nuclear density and the mean distance between baryons is about one Fermi (smaller than one Fermi), where the strong interaction plays an important role. This is the one of necessary conditions for the electric processes of production and oscillation of electron-positron pairs together with ``non-equilibrium'' overcritical electric fields to occur. Under this consideration, the initial radius $R_0$ of the baryon core starting to collapse should be smaller than $30M_0$. However, in this simplified toy model of thin shell collapsing, the surface density of the baryon thin shell is over the nuclear density at the initial radius $R_0=30M_0$. Nevertheless, the necessary condition of baryon cores being at/over the nuclear density should be duly taken into account, when we study the back-reaction in a more realistic model describing the gravitational collapse of neutral stellar cores. 

\begin{figure}[th]
\begin{center}
\includegraphics[width=8cm]{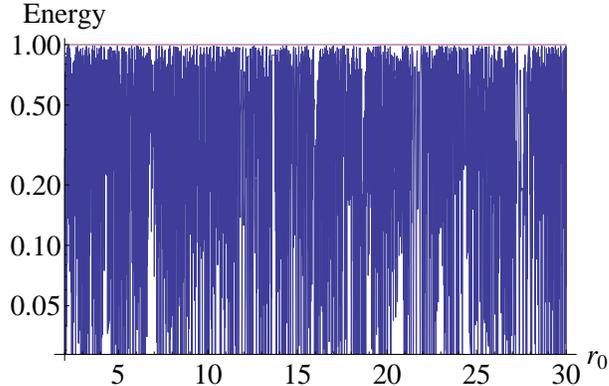}
\end{center}
\caption{In unit of the gravitational energy $M_0^2/(2r_0)$, the gravitational energy (constant red line at $1$) and kinetic energy (fast oscillating lines in blue) and electric energy (fast oscillating lines in white) of the thin shell are plotted as a function of collapsing radius $r_0$.}%
\label{gc_fig3}%
\end{figure}

Using the velocity $\dot r_0 = dr_0/dt_0$ of Eqs.~(\ref{EQUAISRDLC}) and (\ref{coll}), we plot in Fig.~\ref{gc_fig3} the kinetic energy $T(r_0)$ of Eq.~(\ref{EQUY-k}) and the gravitational energy $M_0^2/2r_0$ of the collapsing thin shell as a function of collapsing radius $r_0$. Following the total energy conservation of Eq.~(\ref{EQC}) and $M(r_0)=M_0$, 
\begin{align}
T(r_0)-\tfrac{M_{0}^{2}}{2r_{0}}+\tfrac{Q^{2}_{\rm eff}}{2r_{0}}=0, \label{EQC_1}%
\end{align}
the electric energy $\tfrac{Q^{2}_{\rm eff}}{2r_{0}}$ is given by the difference between gravitational energy and kinetic energy, as shown in Fig.~\ref{gc_fig3}. In the collapse process,
the kinetic energy $T(r_0)$ and electric energy $\tfrac{Q^{2}_{\rm eff}}{2r_{0}}$ are rapidly oscillating, following the ansatz function (\ref{xdecay}) with the frequency $\omega_{\rm osci}$ of microscopic processes. Averaging over these rapid oscillations, we obtain the averaged values of the kinetic energy and electric energy, which are approximately equal to an half of gravitational energy: 
\begin{align}
\langle T(r_0)\rangle \approx  \langle \tfrac{Q^{2}_{\rm eff}}{2r_{0}} \rangle\approx \tfrac{1}{2}\tfrac{M_0^2}{2r_0}. 
\label{EQC_2}%
\end{align}
This implies that the averaged electric energy radiating away from the thin shell is about an half of the gravitational energy gained by the collapsing thin shell in the collapsing process. When the black hole horizon is reached, 
using Eq.~(\ref{EQC}), the irreducible mass of black hole is introduced \cite{RV2002}
\begin{align}
M &= M_{\rm ir} + \tfrac{Q^{2}_{\rm eff}}{2r_{+}},\quad {\rm and}\quad
M_{\rm ir} = M_0- \tfrac{M_0^2}{2r_+} +T(r_+), 
\label{irre}%
\end{align}
where $\tfrac{Q^{2}_{\rm eff}}{2r_{+}}$ is the total electric energy of the thin shell approaching the horizon $r_+$. 
Suppose that the electric energy $\tfrac{Q^{2}_{\rm eff}}{2r_{+}}$ completely radiates away, a black hole is formed with the horizon $r_0\rightarrow r_+=2 M_0$ for $F\equiv f_+(r_0)\rightarrow 0$. In this case,
the total electric energy radiating away from the thin shell is about an half of gravitational energy of the thin shell
\begin{align}
\langle\frac{Q^{2}_{\rm eff}}{2r_{+}}\rangle \approx \frac{1}{2}\left(\frac{M^2_0}{2r_+}\right)=\frac{1}{8}M_0,
\label{radia}
\end{align} 
and the irreducible mass of the formed black hole is about
\begin{align}
M_{\rm ir}
&= M_0- \tfrac{M_0^2}{2r_+} + \langle T(r_+)\rangle 
\approx  \frac{7}{8}M_0, 
\label{irre_1}\\
M_0
&= M_{\rm ir}+\langle\frac{Q^{2}_{\rm eff}}{2r_{+}}\rangle, 
\label{irre_2}%
\end{align}
which implies about $1/8$ of the gravitational energy extracted in gravitational collapses. 
\comment{
In general, suppose that the electric energy $\tfrac{Q^{2}_{\rm eff}}{2r_{+}}$ partially radiates away, the horizon is given by  $r_0\rightarrow r_+\in (GM_0,2 GM_0)$;
} 

\comment{
\begin{figure}[th]
\begin{center}
\includegraphics[width=8cm]{eecollapf.eps}
\end{center}
\caption{In unit of gravitational energy, the electric energy $\tfrac{Q^{2}_{\rm eff}}{2r_{0}}$ radiating away from the thin shell is shown (fast oscillating line in white) as a function of collapsing radius $r_0$.}%
\label{gc_fig4}%
\end{figure}
}

\noindent
{\bf Summary and remarks}
\hskip0.1cm
In this article, on the basis of a simple model for describing the gravitational collapse of a spherical thin-shell capacitor, we analytically study how the gravitational energy gained in collapse converts to the kinetic energy and electric energy, the latter can be radiated away. Using an ansatz function for the effective ``charge-mass-ratio'' (\ref{effx}) to model the microscopic processes that create this electric energy and radiate it away in the Compton scale, we study how the back-reaction of such radiative electric energy on the macroscopic process of gravitational collapse. We find that the rebuilding and radiating of repulsive electric energy cause the collapse process undergoing a sequence of ``on and off'' hopping steps in the microscopic Compton scale. Although such a collapse process is still continuous in the macroscopic scales, it is slowed down as the kinetic energy is reduced and collapsing time is about an order of magnitude larger than that of collapse  process eliminating electric processes. The averaged kinetic and electric energies are the same order, about an half of gravitational energy in collapse.

These results are obtained from an over simplified model for both
macroscopic and microscopic processes. Nevertheless they indicate that apart from an electromagnetic energy radiation, the microscopic processes of electrodynamics have significant back-reaction and effects on gravitational collapsing processes in macroscopic scales. It is thus essential to take into account, 
rather than ignore, electric processes in more realistic models for studying gravitational collapse of neutral stellar core at/over the nuclear density, even though calculations are very complicate. 

To end this article, we would like to mention the relevance of these results to our previous studies of energetic
budget and time duration of Gamma-Ray Bursts (GRBs) as a signal of the final stage of gravitational collapse of massive stellar cores. The total electromagnetic energy extractable from a charged black hole \cite{DR1975,dyado1} (from the collapse of a neutral stellar core \cite{hrx2012}) is a fraction of its mass, which reasonably accounts for the energetic budget of GRBs. In addition, the time duration $T_{90}$ of electromagnetic radiation is about $ 10^{-2}$ second obtained \cite{RSWX1999} by solving hydrodynamical equations with an initial configuration of electro-positron pairs and photons sphere (dyadosphere) around a charged black hole. This time duration scale is elongated to be an order of magnitude larger $\sim 10^{-1}$ second \cite{RVX03c,FRVX2005} by considering both the dynamical formation and hydrodynamical evolution of dyadosphere in a collapsing charged core. The results of this article imply that
due to the back-reaction of the dynamical formation and hydrodynamical evolution of dyadosphere on collapsing neutral stellar cores at or over the nuclear density, the slowing down of gravitational collapsing processes  should elongate this time duration scale by another factor of $10$, i.e., $T_{90}\sim 1$ second that reasonably accounts for the time duration of short GRBs.    
   



\begin{thebibliography}{99}


\bibitem{ob_1975} E.~Olson and M.~Bailyn, Phys.~Rev.~D 12, 3030 (1975), {\it ibid} D 13, (1976) 2204 and D 18, (1978) 2175;\\
N.~K.~Glendenning, ``Compact Stars'' (2000), 
A\&A Library, Springer-Verlag, New York, Chapter 9, pages 340-349;\\ M.~Rotondo, Jorge A.~Rueda, R.~Ruffini and S.-S.~Xue, Phys.~ Lett.~B 701, (2011) 667.

\bibitem{Usov1} Quark stars:  
C.~Alcock, E.~Farhi, A.~Olinto, 
ApJ, vol.~310 (1986) 261;\\ 
V.~V.~Usov, Phys.~Rev.~Lett.~80, (1998) 230;\\
V.~V.~Usov, T.~Harko , K.~S.~Cheng, Astrophys.~J.~620 (2005) 915.\\
Neutron stars: M.~Rotondo, R.~Ruffini and S.-S.~Xue, Int.~J.~Mod.~Phys.~D16 (2007) 1;\\
M.~Rotondo, Jorge A.~Rueda, R.~Ruffini and S.-S.~Xue, Phys.~ Rev.~ C 83, (2011) 045805;\\ 
R.~Belvedere, D.~Pugliese, 
J.~Rueda, R.~Ruffini, S.-S.~Xue, Nuclear Physics A 883 (2012) 1;\\
M.~Rotondo, R.~Ruffini, S.-S.~Xue, and V.~Popov, Int.~J.~ Mod.~ Phys.~ D 20, (2011) 1995;\\
J.~Rueda, R.~Ruffini and S.-S.~Xue, Nucl.~Phys.~A872 (2011) 286.

\bibitem{hrx2012} W.-B.~Han, R.~Ruffini, S.-S.~Xue, Phys.~Rev.~D 86, (2012) 084004.

\bibitem{phreport} R.~Ruffini, G.~V.~Vereshchagin, S.-S.~Xue, Phys.~Rep.~ 487 (2010) 1.

\bibitem{RV2003}
R.~Ruffini and L.~Vitagliano, 
Int.~J.~Mod.~Phys.~D12 (2003) 121. 

\bibitem{Weinberg1972} For the form of these equations, please see, S. Weinberg, `` Gravitation and Cosmology''
ISBN 0-471-92567-5, John Wiley and Sons, 1972.

\bibitem {RVX03a} R.~Ruffini, L.~Vitagliano,  S.-S.~Xue, Phys.~Lett.~
B559 (2003) 12.


\bibitem{I66} W.~Israel, Il Nuovo Cimento B serie 44 (1966) 1, Phys.~Rev.~Lett.~57 (1986) 397;\\ 
V.~De la Cruz and W.~Israel, Il Nuovo Cimento 51A (1967) 744;\\ D.~G.~Boulware Phys.~ Rev.~D 8 (1973)  2363;\\ 
V.~Belinski, M.~Pizzi and A.~Paolino,
Int.~J.~Mod.~Phys.~D 18 (2009) 513;\\
V.~A.~Berezin, V.~A.~Kuzmin and I.~I.~Tkachev,
  Phys.\ Rev.\ D {\bf 36}, 2919 (1987);\\
A.~Aurilia, G.~Denardo, F.~Legovini and E.~Spallucci,
Nucl.\ Phys.\ B {\bf 252}, 523 (1985);\\
S.~Ansoldi,
Class.\ Quant.\ Grav.\ {\bf 19}, 6321 (2002).


\bibitem{crv2002}
C.~Cherubini, R.~Ruffini and L.~Vitagliano, Phys.~Lett.~B545 (2002) 226.

\bibitem{RV2002}
R.~Ruffini and L.~Vitagliano, 
Phys.~Lett.~B545 (2002) 233-237.

\bibitem {RVX03c} R.~Ruffini, L.~Vitagliano,  S.-S.~Xue, Phys.~Lett.~
B573 (2003) 33.

\bibitem {DR1975}
T.~Damour and R.~Ruffini, Phys.~Rev.~Lett.~35 (1975) 463. 

\bibitem{dyado1} G.~Preparata, R.~Ruffini, S.-S.~Xue, Astron.Astrophys., 338, L87-L90 (1998) arXiv:astro-ph/9810182v1, J.~Korean Phys.~Soc.~42: S99-S104, 2003, arXiv:astro-ph/0204080v1;\\
R.~Ruffini, S.-S.~Xue, a review in AIP Conf.~Proc., 1059 (2008) 72  (arXiv:0810.1438);\\
C.~Cherubini, A.~Geralico, J.~A.~Rueda H., and R.~ Ruffini, Phys.~Rev.~D 79, (2009) 124002.

\bibitem {RSWX1999}
R.~Ruffini, J.~D.~Salmonson, J.~R.~Wilson, and S.-S.Xue, A\&A 350 (1999) 334,
{\it ibid}, 359 (2000) 855.

\bibitem {FRVX2005} F.~Fraschetti, R.~Ruffini, L.~Vitagliano,  S.-S.~Xue,  	Int.~J.~Mod.~Phys.~ D14 (2005) 131,  Nuovo Cim.~B121 (2006) 1477, (the Proceedings of the ``Swift and GRBs: Unveiling the Relativistic Universe'', in Venice (Italy), June 5-9, 2006).

\end{thebibliography}
\end{document}